\documentclass[prl,aps,reprint,groupedaddress,showkeys]{revtex4-1}

\usepackage[utf8]{inputenc}

\usepackage{fullpage}
\usepackage[T1]{fontenc}
\usepackage{verbatim}
\usepackage{amssymb,amsmath}
\usepackage{comment,xcolor}
\usepackage[colorlinks,linktocpage,linkcolor=blue,citecolor=red,urlcolor=red]{hyperref}

\def\be{\begin{equation}} 
\def\ee{\end{equation}}

\allowdisplaybreaks

\begin{document}

\title{Smooth Scale Separation and Holography}

\author{Aviral Aggarwal$^1$}
\author{Niccol\`o Cribiori$^2$} 
\author{Fotis Farakos$^3$} 
\author{Alexandros Zarafonitis$^3$}

\affiliation{}
\affiliation{$^1$ Department of Theoretical Physics and Astrophysics, 
Faculty of Science, Masaryk University, 611 37 Brno, Czech Republic} 
\affiliation{$^2$ Inst.~for Theoretical Physics, KU Leuven, Celestijnenlaan 200D, B-3001 Leuven, Belgium}
\affiliation{$^3$ Physics Division, National Technical University of Athens
15780 Zografou Campus, Athens, Greece} 

\begin{abstract}

We construct candidate supersymmetric scale-separated AdS$_4$ vacua of massive type IIA string theory on an orientifold of a smooth Calabi--Yau threefold, a free quotient of $T^6$ with non-abelian point group $D_4$. The internal space has no orbifold singularities or O-plane intersections, and the effective theory has only five chiral multiplets. The model features nonzero extremal cubic couplings violating a recent holographic consistency condition. One interpretation is that a leading effective theory may be a valid low-energy approximation without being a complete holographic description.

\end{abstract}

\maketitle

\section{Introduction}

A compactification of string theory describing the four-dimensional world requires \emph{scale separation}: the lowest tower of Kaluza--Klein (KK) modes must lie well above the curvature scale of the vacuum \cite{Coudarchet:2023mfs}. The KK scale then sets an upper bound on the cut-off of the lower-dimensional low-energy effective theory (EFT), but the omitted modes can still affect observables below that scale. 
Whether the required hierarchy between the KK scale and the curvature scale of the vacuum can be achieved dynamically in a controlled setup remains open. No-go theorems and general arguments exclude it in broad classes of compactifications \cite{Gautason:2015tig,Lust:2019zwm,Cribiori:2022trc}, while candidate constructions appear to achieve it \cite{DeWolfe:2005uu,Camara:2005dc,Farakos:2020phe,Cribiori:2021djm,Tringas:2025bwe}, albeit within approximations such as smeared orientifold planes.

Restricting to supersymmetric anti-de Sitter (AdS) vacua simplifies the problem. 
Besides smearing, all explicit candidates are built on singular orbifolds, with blow-up modes and possible O-plane intersections that further complicate their microscopic description. 
Holographically, parametric scale separation requires a family of conformal field theories (CFTs) with large central charge $c$ and a parametrically large gap in the single-trace spectrum above finitely many light operators. 
The CFT duals of candidate scale-separated vacua remain unknown, and their existence would be in tension with a conjectured upper bound on the spin-two KK gap \cite{Collins:2022nux}. 
Recently, holographic consistency conditions for scale-separated vacua have been proposed in terms of the masses and cubic couplings of the light fields in the lower-dimensional EFT, with the KK and string towers truncated \cite{Bobev:2025yxp,Revello:2026yla}. 
This raises a central question: How much can the leading lower-dimensional EFT alone tell us about the holographic consistency of the vacuum?

In this Letter we report two results. 
First, we construct minimal candidate scale-separated AdS$_4$ vacua. 
These vacua arise from compactifying massive type IIA string theory on an orientifold of the Calabi--Yau threefold $X=T^6/\Gamma$, where $\Gamma$ acts freely and has point group $D_4$, the dihedral group of order eight \cite{Catanese:2018}. The quotient is smooth and has no blow-up modes, the O6-planes wrap three disjoint cycles, and the internal metric is explicit and flat in the smeared approximation. The four-dimensional EFT contains only five chiral multiplets, and all saxions are stabilized at large volume and weak coupling. The model is thus a clean laboratory for the open questions of scale separation 
and orientifold backreaction \cite{Junghans:2020acz,Marchesano:2020qvg}.

Second, we find nonzero extremal cubic couplings in the leading four-dimensional 
EFT, in violation of the holographic constraint proposed in \cite{Bobev:2025yxp}, 
and extended in \cite{Revello:2026yla}.  
This discrepancy admits two possible interpretations. 
Since the hierarchy between the KK and AdS scales diverges only in the large-flux limit, corrections from the omitted towers could allow the full compactification to evade the constraint even when the leading EFT properly describes the low-energy physics. Alternatively, the constraint could signal a genuine tension between this class of scale-separated vacua and holography.

\section{An explicit Calabi--Yau threefold}

The smooth internal space we work with is the toroidal orbifold $X=T^6/\Gamma$, where $\Gamma$ acts freely and has non-abelian point group $D_4$ \cite{Oguiso:1999xq,Fischer:2013qza}. The free action gives $\chi(X)=\chi(T^6)/|\Gamma|=0$ and, since $\chi(X)=2(h^{1,1}-h^{2,1})$ for a Calabi--Yau threefold, every such free Calabi--Yau quotient has $h^{1,1}=h^{2,1}$. The $D_4$ family realizes $(h^{1,1},h^{2,1})=(2,2)$ and thus attains the smallest value of $h^{1,1}+h^{2,1}$ tabulated for the 469 symmetric toroidal orbifolds preserving one quarter of the supersymmetry of type~II string theory on $T^6$ \cite{Fischer:2012qj,Fischer:2013qza}. It also minimizes both Hodge numbers among Calabi--Yau threefolds of type~A, namely free quotients of abelian threefolds \cite[Theorem~0.1]{Oguiso:1999xq} \footnote{Braun constructed smooth Calabi--Yau threefolds with Hodge pair $(1,1)$ as free quotients of a hypersurface in a toric fourfold \cite{Braun:2011hd}. However, their finite fundamental groups rule out a torus cover.}.

On $T^6$ we choose periodic real coordinates $x_i\sim x_i+1$ and $y_i\sim y_i+1$ and combine them into complex coordinates
\begin{equation}
z_1=x_1+\tau y_1,\quad
z_2=x_2+\tau y_2,\quad
z_3=x_3+\tau' y_3,
\end{equation}
with $\tau,\tau'$ in the upper half-plane. The group $\Gamma$ acting on this torus is generated by
\begin{equation}
\begin{aligned}
r:(z_1,z_2,z_3)&\mapsto \left(z_2,\,-z_1,\,z_3+\tfrac14\right),\\
s:(z_1,z_2,z_3)&\mapsto \left(z_1+\tfrac12,\,-z_2+\tfrac\tau2,\,-z_3\right).
\end{aligned}
\end{equation}
Omitting the fractional translations gives two linear transformations that generate the dihedral group $D_4$. On $T^6$, the affine maps satisfy $r^4=s^2=1$ and $(rs)^2=w$, where $w$ is the order-two translation
\begin{equation}
w:(z_1,z_2,z_3)\mapsto
\left(z_1+\tfrac{1+\tau}{2},\,z_2+\tfrac{1+\tau}{2},\,z_3\right).
\end{equation}
Since $w$ commutes with both generators, $\langle w\rangle\simeq\mathbb Z_2$ is a normal subgroup of $\Gamma$.
The quotient can therefore be taken in two steps. First, we form $T'=T^6/\langle w\rangle$. This amounts to adding the period $\bigl(\tfrac{1+\tau}{2},\tfrac{1+\tau}{2},0\bigr)$ to the lattice, so $T'$ is again a complex torus. On $T'$, $w$ acts trivially, so the induced maps $r,s$ satisfy the remaining dihedral relation $(rs)^2=1$. Since their linear parts already generate $D_4$, it follows that $\Gamma/\langle w\rangle\simeq D_4$ and hence $|\Gamma|=16$. Second, we quotient $T'$ by this $D_4$ action to obtain
\begin{equation}
X=T^6/\Gamma=T'/D_4.
\end{equation}
The chosen fractional translations make the $D_4$ action on $T'$ free \cite[Theorem~0.1]{Catanese:2018}, so $X$ has no orbifold singularities.
The parameters $\tau,\tau'$ give local coordinates on the complex-structure moduli space of $X$. In view of the orientifold projection, we restrict to $\tau=e^{i\theta}$ and $\tau'=\frac12+i\nu$, with $0<\theta<\pi$ and $\nu>0$. Together with the four-dimensional dilaton, $\theta$ and $\nu$ parametrize the saxions of the complex-structure sector.

The holomorphic three-form
\begin{equation}
\Omega=-\tau^{-1}\,dz_1\wedge dz_2\wedge dz_3
\end{equation}
is invariant under $\Gamma$; its phase is fixed so that the orientifold involution introduced below sends $\Omega$ to $\overline{\Omega}$. It descends to a nowhere-vanishing holomorphic three-form on $X$, which trivializes the canonical bundle. The $r$-invariant holomorphic one- and two-forms are spanned by $dz_3$ and $dz_1\wedge dz_2$, respectively. Both are odd under $s$, so $h^{1,0}(X)=h^{2,0}(X)=0$. Since $X$ is K\"ahler, these properties make it a Calabi--Yau threefold. Its flat metric has trivial restricted holonomy and full holonomy $D_4\subset\mathrm{SU}(3)$, generated by the linear parts of $r$ and $s$ \footnote{For a smooth flat torus quotient, the linear parts of the quotient transformations form the full holonomy group \cite[Sec.~6, Eq.~(7) and Remark~4]{Derdzinski_2025}.  In our case, translations have trivial linear part, while the linear parts of $r$ and $s$ generate $D_4$. Both are unitary and have determinant one, so they preserve the flat metric and $\Omega$, realizing $D_4$ as a subgroup of $\mathrm{SU}(3)$.}.

The invariant K\"ahler form is
\begin{equation}
J=v^1{\cal D}_1+v^2{\cal D}_2,
\end{equation}
where $v^1,v^2>0$ are the two real K\"ahler moduli and 
\begin{equation}
\begin{aligned}
{\cal D}_1 &= 2\,(dx_1\wedge dy_1+dx_2\wedge dy_2),\\
{\cal D}_2 &= 4\,dx_3\wedge dy_3
\end{aligned}
\end{equation}
represent a primitive integral basis of $H^2(X,\mathbb Z)$ modulo torsion. The flat metric corresponding to such K\"ahler form is
\begin{equation}
ds^2=\frac{2v^1}{{\rm Im}\,\tau}\bigl(|dz_1|^2+|dz_2|^2\bigr)
+\frac{4v^2}{{\rm Im}\,\tau'}\,|dz_3|^2.
\end{equation}
Up to permutations of the indices, the only nonzero triple intersection number is $\kappa_{112}=\int_X{\cal D}_1^2\wedge{\cal D}_2=2$. The volume is therefore ${\cal V}=\frac16\int_X J^3=(v^1)^2v^2$, while the dual integral basis of $H^4(X,\mathbb Z)$ modulo torsion is $\widetilde{\cal D}^1=\frac12{\cal D}_1\wedge{\cal D}_2$, $\widetilde{\cal D}^2=\frac12{\cal D}_1^2$.

\section{Orientifold}

On the covering torus $T^6$ we choose the antiholomorphic involution
\begin{equation}
\sigma:(z_1,z_2,z_3)\mapsto(-\tau\bar z_1,\,-\tau\bar z_2,\,\bar z_3),
\end{equation}
which acts separately on the three elliptic factors \cite{Doran:2014qna}. With this choice, fluxes alone can cancel the O6-plane charge in the smeared Bianchi identity for $F_2$, without introducing D6-branes, as we show in the next section.
Since $|\tau|=1$ and ${\rm Re}\,\tau'=\frac12$, $\sigma$ preserves the lattice, acting on the real coordinates as $(x_i,y_i)\mapsto(-y_i,-x_i)$ for $i=1,2$ and $(x_3,y_3)\mapsto(x_3+y_3,-y_3)$. From $\sigma r\sigma=r$ and $\sigma s\sigma=ws$, we get $\sigma\Gamma\sigma=\Gamma$, so $\sigma$ maps $\Gamma$-orbits to $\Gamma$-orbits and descends to an involution of $X$. Furthermore, we have $\sigma^*\Omega=\overline\Omega$ and $\sigma^*J=-J$, as required by \cite{Grimm:2004ua}.

We consider O6$^-$-planes filling four-dimensional spacetime and wrapping the fixed locus $\Sigma_{\rm O6}$ of $\sigma$ on $X$. A point of $X$ is fixed by $\sigma$ if and only if its lifts satisfy $\sigma z=gz$ for some $g\in\Gamma$. This happens for eight elements $g$, each giving a flat three-torus fixed by $g^{-1}\sigma$. These eight tori form three $\Gamma$-orbits, so $\Sigma_{\rm O6}$  has three connected components, represented by the fixed loci of $\sigma$, $r^2\sigma$ and $(rs)^{-1}\sigma$. Each component is special Lagrangian, since both $J$ and ${\rm Im}\,\Omega$ vanish on the fixed locus. These O6-planes never (self-)intersect although they wrap different homology classes: if a point satisfied both $\sigma z=gz$ and $\sigma z=hz$, then $h^{-1}gz=z$, so $g=h$ since $\Gamma$ acts freely. The eight tori are therefore disjoint and are permuted by $\Gamma$, so their images in $X$ form the three embedded components of $\Sigma_{\rm O6}$. This is an advantage of freely acting orbifolds, since O-plane intersections occur only at fixed points of the orbifold group \cite{Junghans:2023yue}.

To describe forms of definite parity, it is convenient to introduce the one-forms
\begin{equation}
\begin{aligned}
e^\pm_i &= dx_i \mp dy_i \quad (i=1,2),\\ 
e^+_3 &= 2\,dx_3 + dy_3 , \qquad e^-_3 = dy_3 ,
\end{aligned}
\end{equation}
which satisfy $\sigma^*e^\pm_i=\pm e^\pm_i$.
A primitive integral symplectic basis of $H^3(X,\mathbb Z)$ modulo torsion, adapted to the orientifold, consists of the even three-forms 
\begin{equation}
\begin{aligned} 
& {\cal A}_0 = \big(e^+_1 \wedge e^+_2 + e^-_1 \wedge e^-_2\big) \wedge e^+_3, 
\\
& {\cal A}_1 = \big(e^+_1 \wedge e^+_2 - e^-_1 \wedge e^-_2\big) \wedge e^+_3 , 
\\ 
& {\cal A}_2 = - \big(e^+_1 \wedge e^-_2 + e^-_1 \wedge e^+_2 \big) \wedge e^-_3 , 
\end{aligned}
\end{equation}
and the odd three-forms
\begin{equation}
\begin{aligned}
& {\cal B}^0 = - \big(e^+_1 \wedge e^+_2 + e^-_1 \wedge e^-_2\big) \wedge e^-_3 , 
\\ 
& {\cal B}^1 = \big(e^+_1 \wedge e^+_2 - e^-_1 \wedge e^-_2\big) \wedge e^-_3  , 
\\ 
& {\cal B}^2 = \big(e^+_1 \wedge e^-_2 + e^-_1 \wedge e^+_2\big) \wedge e^+_3 , 
\end{aligned}
\label{CalB}
\end{equation}
with $\int_X {\cal A}_I \wedge {\cal B}^J = \delta_I^J$ and 
$\int_X{\cal A}_I\wedge{\cal A}_J=0=\int_X{\cal B}^I\wedge{\cal B}^J$. 
In this basis, 
\begin{equation}
\begin{aligned}
{\rm Re}\, \Omega &= -\frac{\cos\theta}{4}\, {\cal A}_0 + \frac{1}{4}\, {\cal A}_1 + \frac{\nu\sin\theta}{2}\, {\cal A}_2 , 
\\
{\rm Im}\, \Omega &= \frac{\nu\cos\theta}{2}\, {\cal B}^0 + \frac{\nu}{2}\, {\cal B}^1 + \frac{\sin\theta}{4}\, {\cal B}^2 ,
\end{aligned}
\label{eq:Omegaexpansion}
\end{equation}
and the Poincar\'e dual of $\Sigma_{\rm O6}$, with each component oriented by ${\rm Re}\,\Omega$, is 
\begin{equation}
{\rm PD}[\Sigma_{\rm O6}]=2\,({\cal B}^1+{\cal B}^2).
\end{equation}
Proofs of the integrality statements and of the structure of the O6 locus are given in the Appendix.

\section{Fluxes, tadpole and solution}

We now consider massive type IIA string theory on the orientifold of the smooth Calabi--Yau $X$ we have just constructed.
We normalize the RR fluxes $F_0,F_2,F_4$ and the NSNS flux $H_3$ so that their background flux quanta are integers in our chosen bases on $X$ \footnote{A more precise classification of RR fluxes is by (twisted, and for orientifolds KR-) K-theory \cite{Minasian:1997mm,Witten:1998cd}, and their quantization receives curvature corrections, such as the $p_1$-shift of $F_4$ known from M-theory \cite{Witten:1996md}. On our flat $X$ these corrections vanish, and the remaining refinements are discrete: torsion fluxes, which we do not turn on, and parity conditions from the orientifold, such as even $m_0$ for O6$^-$-planes \cite{Bergman:2001rp}. In the smeared vacuum the $B$-field axions vanish, so the four-form component of the Page flux $e^{-B_2}\wedge F$ \cite{Marolf:2000cb} has the same periods as $F_4$.}. 
Compatibility with the orientifold involution requires $F_0,F_4$ to be even and $F_2,H_3$ to be odd under $\sigma$ \cite{Grimm:2004ua}. In the smeared approximation we set $F_2=0$ and expand the remaining fluxes in the primitive bases introduced above,
\begin{equation}
F_0=m_0,\qquad F_4=e_a\,\widetilde{\cal D}^a,\qquad H_3=h_I\,{\cal B}^I ,
\end{equation}
where $m_0,e_a,h_I\in\mathbb Z$, with $a=1,2$ and $I=0,1,2$.

On the covering space $X$, a bulk D6-brane and its image carry two units of charge in total, and an O6$^-$-plane carries $-4$ units. On the orientifold quotient $X/\sigma$, the corresponding charges are $1$ and $-2$. In the absence of D6-branes, the Bianchi identity for $F_2$ reads $dF_2=F_0H_3-4\,\delta_3(\Sigma_{\rm O6})$, where $\delta_3(\Sigma_{\rm O6})$ is the current of the O6 locus; in the smeared approximation it is replaced by its harmonic representative ${\rm PD}[\Sigma_{\rm O6}]$. In cohomology, the Bianchi identity becomes the tadpole condition
\begin{equation}
m_0\,H_3=4\,{\rm PD}[\Sigma_{\rm O6}]=8\,({\cal B}^1+{\cal B}^2) ,
\end{equation}
that is, $m_0h_0=0$ and $m_0h_1=m_0h_2=8$.
For O6$^-$-planes, $m_0$ must be even \cite{Bergman:2001rp}, leaving the possible flux choices $(m_0;h_0,h_1,h_2)\in\{\pm(2;0,4,4),\,\pm(4;0,2,2),\,\pm(8;0,1,1)\}$.
The tadpole is thus canceled by fluxes alone, without D6-branes. The remaining Bianchi identities are trivially satisfied, so $F_4$ is unconstrained.

These data specify a supersymmetric AdS$_4$ solution of the ten-dimensional equations of motion in the smeared approximation, with constant dilaton and no warping. For a Calabi--Yau orientifold with $F_2=0$, the supersymmetry conditions require $H_3\propto{\rm Im}\,\Omega$ and $F_4\propto J\wedge J$ \cite{Lust:2004ig,DeWolfe:2005uu}. Since the tadpole sets $H_3\propto{\cal B}^1+{\cal B}^2$, the first condition fixes the complex structure to $\theta=\pi/2$ and $\nu=\tfrac12$. Using $J\wedge J=4v^1v^2\,\widetilde{\cal D}^1+2(v^1)^2\,\widetilde{\cal D}^2$, the second gives $v^1/v^2=2e_2/e_1$, with both $e_1$ and $e_2$ opposite in sign to $m_0$ on the supersymmetric branch. At fixed $m_0$ and $h_I$, the overall volume and the dilaton are then determined by the magnitudes of the $e_a$. Rescaling $e_a\to n\,e_a$ with $n\to\infty$, the vacuum sits at large volume and weak coupling, ${\cal V}\sim n^{3/2}$ and $g_s\sim n^{-3/4}$, and exhibits parametric scale separation, $L_{\rm KK}/L_{\rm AdS}\sim n^{-1/2}$, as in \cite{DeWolfe:2005uu}.

\section{Effective theory and holography}

The orientifold has $h^{1,1}_+=0$ and $h^{1,1}_-=2$, while $H^3(X)$ splits into even and odd subspaces, spanned by ${\cal A}_I$ and ${\cal B}^I$, each of dimension $h^{2,1}+1=3$. The four-dimensional $\mathcal N=1$ theory therefore has no vector multiplets and $h^{1,1}_-+h^{2,1}+1=5$ chiral multiplets \cite{Grimm:2004ua}. In the K\"ahler sector, the $B$-field $B_2=b^a{\cal D}_a$ provides two axions, which pair with the K\"ahler moduli into $T^a=b^a+iv^a$.
In the complex-structure sector, the RR three-form $C_3=\xi^I{\cal A}_I$ provides three axions, which pair with $\theta$, $\nu$ and the four-dimensional dilaton $e^{-D}=e^{-\phi}\sqrt{\cal V}$ into ${\cal N}^I=\xi^I+iu^I$. The saxions $u^I$ are defined by $u^I{\cal A}_I=2\sqrt 2 \,e^{-D}{\rm Re}\,\Omega/(i\int_X\Omega\wedge\overline\Omega)^{1/2}$, where $i\int_X\Omega\wedge\overline\Omega=\frac12\nu\sin^2\theta$. 
Using the expansion of ${\rm Re}\,\Omega$ in \eqref{eq:Omegaexpansion}, we find $u^1=\frac{e^{-D}}{\sqrt{\nu}\,\sin\theta}$, $u^0=-\cos\theta\ u^1$ and $u^2=2\nu\sin\theta\ u^1$.
In terms of these fields, the K\"ahler potential and the superpotential are
\begin{equation}
\begin{aligned}
K&=-\log(8{\cal V})+4D, \\
W&=e_aT^a-m_0\,(T^1)^2T^2+h_I{\cal N}^I ,
\end{aligned}
\end{equation}
where ${\cal V}=(v^1)^2v^2$, $e^{-4D}=\tfrac14\big((u^1)^2-(u^0)^2\big)(u^2)^2$, and the tadpole sets $h_I{\cal N}^I=\frac{8}{m_0}({\cal N}^1+{\cal N}^2)$. 
Since $K$ depends only on the saxions and $W$ does not depend on ${\cal N}^0$ and ${\cal N}^1-{\cal N}^2$, the axions $\xi^0$ and $\xi^1-\xi^2$ are perturbatively flat. They take values in compact spaces and may be lifted by non-perturbative effects.

Using $m^2L_{\rm AdS}^2=\Delta(\Delta-3)$, we obtain the leading dimensions of the operators dual to the ten real scalars. Grouped by chiral multiplet (axion, saxion), they are
\begin{equation}
(3,2),\quad (3,2),\quad (5,6),\quad (5,6),\quad (11,10). 
\end{equation}
These dimensions are independent of fluxes, while the central charge of the putative CFT scales as $c\sim n^{9/2}$ in the large-flux limit.

We now test the holographic constraint of \cite{Bobev:2025yxp}. 
A cubic coupling $c'_{ijk}$ of three bulk scalars is called extremal when the dimension of one dual operator equals the sum of the other two, $\Delta_i=\Delta_j+\Delta_k$. Under the large-$c$ assumptions of \cite{Bobev:2025yxp}, if this relation holds at tree level, the corresponding on-shell coupling $c'_{ijk}$ must vanish at that order.
Our spectrum contains four extremal triples of dimensions.  
For $(10;5,5)$ and $(11;5,6)$, the relevant cubic interactions follow from the untwisted sector of the toroidal orbifold $T^6/(\mathbb{Z}_3 \times \mathbb{Z}_3)$ \cite{DeWolfe:2005uu} by identifying two K\"ahler multiplets and rescaling the third to match our normalization. The on-shell couplings therefore vanish by the same cancellation found in \cite{Bobev:2025yxp}.
The remaining two triples involve the complex-structure multiplets and thus we expect non-vanishing extremal couplings \cite{Revello:2026yla}. 
Indeed, an explicit calculation gives $c'_{5,2,3}=12\sqrt{3/13}$ and $c'_{6,3,3}=18\sqrt{3/13}$ in the normalization of \cite{Bobev:2025yxp}. 
Our model thus violates the holographic constraint in the leading four-dimensional EFT.

What this violation implies for the full vacuum depends on how the tree-level relation $\Delta_i=\Delta_j+\Delta_k$ is corrected at large $c$. The four-dimensional EFT retains only the light fields below $M_{\rm KK}$, with the KK and string towers truncated. Loops of the retained fields give corrections of order $(M_P L_{\rm AdS})^{-2}\sim 1/c$ to the dimensions \cite{Bobev:2025yxp}. In this setup, however, the masses of the omitted towers vary with $c$ in AdS units, so integrating them out could give a larger correction to $\delta_{ijk}=\Delta_i-\Delta_j-\Delta_k$.
To see how the constraint depends on the corrections to the dimensions, suppose that the first nonzero contribution scales as $c^{-\alpha}$. In the normalization of \cite{Bobev:2025yxp}, a single-particle operator has a two-point function of order $1/c$. If it contains a relative double-trace admixture of order $c^{-\beta}$, preserving the leading $1/c$ behavior of the two-point function requires $\beta\geq0$.
Now consider the three-point function. Near extremality, the bulk three-point integral is proportional to $1/\delta_{ijk}$. For a nonzero leading coupling $c'_{ijk}$ and $|\delta_{ijk}|\sim c^{-\alpha}$, this enhances the three-point coefficient from order $c^{-2}$ to order $c^{\alpha-2}$. In the CFT three-point function, the overall normalization $c^{-3/2}$ multiplies the double-trace mixing coefficient $c^{-\beta}$, giving a contribution of order $c^{-3/2-\beta}$. Matching the two powers gives
\begin{equation}
 \beta=\frac12-\alpha.
\end{equation}
For $\alpha=1$, this requires $\beta=-\tfrac12$, in conflict with $\beta \geq 0$ argued for before. At $\alpha=\tfrac12$ the bound is saturated, while for $0<\alpha<\tfrac12$, the double-trace contribution vanishes relative to the single-trace term at large $c$, so the contradiction no longer arises.

The flux dependence of the lower-dimensional EFT can give corrections larger than $1/c$. For example, the $(\alpha')^3$ correction studied in the mirror description of DGKT vacua \cite{Plauschinn:2022ztd} shifts the dimension of the coupled $\Delta=5$ mode while leaving the $\Delta=2,3$ modes unchanged. Its contribution to $\Delta_5-\Delta_2-\Delta_3$ is proportional to $\chi(X)/{\cal V}$. For $\chi(X)\ne0$, this term scales as $c^{-1/3}$ along the isotropic large-flux family, since ${\cal V}\sim c^{1/3}$. For our smooth quotient it vanishes because $\chi(X)=0$. 
The KK and string oscillator towers may also contribute to the dimensions of light operators. In our model, $M_{\rm KK}/M_s\sim{\cal V}^{-1/6}$, so
\begin{equation}
\begin{aligned}
 r_{\rm KK}&\equiv(M_{\rm KK}L_{\rm AdS})^{-2}\sim n^{-1}\sim c^{-2/9},\\
 r_s&\equiv(M_sL_{\rm AdS})^{-2}\sim n^{-3/2}\sim c^{-1/3}.
\end{aligned}
\end{equation}
A leading correction to $\delta_{ijk}$ proportional to $r_{\rm KK}$, $r_{\rm KK}^2$ or $r_s$ would give $\alpha=2/9$, $4/9$ or $1/3$, respectively, all below the threshold $\alpha=\tfrac12$. It remains to determine whether corrections of these orders arise and, if so, whether they shift either $\Delta_5-\Delta_2-\Delta_3$ or $\Delta_6-2\Delta_3$ away from zero 
fast enough to evade \cite{Bobev:2025yxp}. 
More generally, quantum corrections in minimally supersymmetric setups are still only partially known. 
Note that not any descent of towers would work: 
it has to be sufficiently fast in Planck units, a condition reminiscent of the distance conjecture \cite{Ooguri:2006in,Lust:2019zwm}.

\section{Discussion}

We have constructed candidate scale-separated supersymmetric AdS$_4$ vacua in the smeared approximation of massive type IIA supergravity on an orientifold of a smooth toroidal quotient with non-abelian point group $D_4$ and $(h^{1,1},h^{2,1})=(2,2)$. The internal metric is explicit and flat, there are no blow-up modes, and the O6-planes wrap three disjoint cycles in distinct homology classes. The four-dimensional EFT contains only five chiral multiplets and has nonzero extremal cubic couplings involving the complex-structure sector.

Our construction points to a new path to supersymmetric scale separation through toroidal orbifolds with \emph{non-abelian} orbifold groups and very few moduli. Here we have used a freely acting $D_4$ quotient with $h^{1,1}=h^{2,1}=2$ and no blow-up moduli. Another possibility, among the examples classified in \cite{Fischer:2013qza}, is to start from an orbifold whose resolution gives a rigid Calabi--Yau threefold with few K\"ahler moduli. For example, resolving the $\Delta(27)$ orbifold gives $(h^{1,1},h^{2,1})=(4,0)$. Our preliminary checks indicate that O6-planes can be included consistently and that the resulting orientifold is compatible with scale separation. 
Reducing further on $S^1$ with additional fluxes and sources can lead to scale-separated AdS$_3$ vacua \cite{AAAZ}. 
AdS$_3$ constructions with $A_4\times\mathbb Z_2$ and $D_4\times\mathbb Z_2$ were studied in \cite{Revello:2026eqp}.

The violation of the holographic constraint in the leading EFT admits two interpretations. Corrections from the omitted towers may change the large-$c$ scaling of the extremal combinations and allow the full compactification to evade the constraint. On this interpretation, a purely bottom-up holographic description based on the truncated EFT would be incomplete. Even the near-extremal three-point functions of light operators would require input from the ultraviolet completion, although the EFT remains reliable below the KK scale at fixed flux. 
In other words, the finite-field EFT may be a valid low-energy description of an AdS vacuum without being a complete holographic description of that vacuum. 
Alternatively, if the constraint remains applicable, the nonzero couplings signal a genuine tension between this class of scale-separated vacua and holography.

\vspace{.2cm}

\begin{acknowledgments}
\emph{ We thank V.~Van Hemelryck for discussions. 
A.A.~is supported by the Czech Science Foundation
GACR grant “On and Off-shell Superspace Representations” (GA26-23375S) and thanks the Physics Department of NTUA for the hospitality during the completion of this work. 
The work of N.C.~is supported by the Research Foundation Flanders (FWO grant 1259125N).}
\end{acknowledgments}

\appendix

\section{Appendix: CY3 properties}

We prove the statements of the main text on the group $\Gamma$, the primitive bases ${\cal D}_a$, $\widetilde{\cal D}^a$ and $({\cal A}_I,{\cal B}^I)$, and the integral cohomology of $X$ modulo torsion.

\subsection*{A. Conventions and the group $\Gamma$}

We use real coordinates $q=(x_1,y_1,x_2,y_2,x_3,y_3)$ on $T^6=\mathbb R^6/\mathbb Z^6$, each of period one. In these coordinates, the generators $r,s$ of $\Gamma$, the translation $w$ and the orientifold involution $\sigma$ act as
\begin{equation}
\begin{aligned}
 r&:\big(x_2,\,y_2,\,-x_1,\,-y_1,\,x_3+\tfrac14,\,y_3\big),  \\ 
 s&:\big(x_1+\tfrac12,\,y_1,\,-x_2,\,-y_2+\tfrac12,\,-x_3,\,-y_3\big),\\
w&:\big(x_1+\tfrac12,\,y_1+\tfrac12,\,x_2+\tfrac12,\,y_2+\tfrac12,\,x_3,\,y_3\big), \\ 
 \sigma&:\big(-y_1,\,-x_1,\,-y_2,\,-x_2,\,x_3+y_3,\,-y_3\big).
\end{aligned}
\end{equation}
Modulo $\mathbb Z^6$, these maps satisfy $r^4=s^2=\sigma^2=1$, $(rs)^2=w$, $\sigma r\sigma=r$ and $\sigma s\sigma=ws$. Since $w$ is central of order two and $sr=r^{-1}sw$, every element of $\Gamma$ can be written as $r^as^bw^c$, with $0\leq a\leq 3$ and $b,c\in\{0,1\}$. These sixteen elements are distinct because the eight $r^as^b$ have distinct linear parts and $w$ is a nontrivial translation. Thus $|\Gamma|=16$, with presentation
\begin{equation}
 \Gamma=\big\langle r,s\,\big|\,r^4,\,s^2,\,(rs)^4,\,[r,(rs)^2],\,[s,(rs)^2]\big\rangle ,
\end{equation}
and $\Gamma/\langle w\rangle\simeq D_4$.
For the quotient map $\pi:T^6\to X$, we have $\int_X\omega=\frac1{16}\int_{T^6}\pi^*\omega$ for any six-form $\omega$. We orient $T^6$ by $d^6x=dx_1\wedge dy_1\wedge dx_2\wedge dy_2\wedge dx_3\wedge dy_3$, with $\int_{T^6}d^6x=1$. We use periodic parameters $t_i\in\mathbb R/\mathbb Z$ for the cycles below.

\subsection*{B. Primitive integral basis: degrees two and four}

The $\Gamma$-invariant constant two-forms on $T^6$ are spanned by
$\omega_1=dx_1\wedge dy_1+dx_2\wedge dy_2$ and $\omega_2=dx_3\wedge dy_3$.
We now show that ${\cal D}_1=2\omega_1$ and ${\cal D}_2=4\omega_2$ form a primitive integral basis of $H^2(X,\mathbb Z)$ modulo torsion.

\paragraph{Necessity.} The maps
\begin{equation}
\begin{aligned}
 C_1(t_1,t_2)&=\big(t_1+\tfrac{t_2}2,\,\tfrac{t_2}2,\,
                  \tfrac{t_2}2,\,\tfrac{t_2}2,\,0,\,0\big),\\
 C_2(t_1,t_2)&=\big(0,\,0,\,0,\,0,\,\tfrac{t_1}4,\,t_2\big)
\end{aligned}
\end{equation}
define two-cycles in $X$, since $C_1(t_1,t_2+1)=w\,C_1(t_1,t_2)$ and $C_2(t_1+1,t_2)=r\,C_2(t_1,t_2)$, while the other unit shifts are lattice translations. Their periods satisfy $\int_{C_a}{\cal D}_b=\delta_{ab}$. Thus the periods of $n\omega_1+m\omega_2$ on $C_1$ and $C_2$ are $n/2$ and $m/4$, respectively, so integrality requires $n\in2\mathbb Z$ and $m\in4\mathbb Z$.

\paragraph{Sufficiency.} To show that these conditions suffice, we construct line bundles on $X$ with first Chern classes $[{\cal D}_1]$ and $[{\cal D}_2]$. For $n,m\in\mathbb Z$, consider the connection $d-2\pi iA$ on $\mathbb R^6\times\mathbb C$, with $A=n(x_1dy_1+x_2dy_2)+m\,x_3dy_3$. Under a lattice shift $q\to q+\ell$, the potential changes by $A(q+\ell)=A(q)+df_\ell$, where
\begin{equation}
 f_\ell(q)=n(\ell_{x_1}y_1+\ell_{x_2}y_2)+m\,\ell_{x_3}y_3,
 \quad \ell\in\mathbb Z^6 .
\end{equation}
To compensate this shift, we let the lattice act on the fiber coordinate $\psi$ by the phase $e^{2\pi i f_\ell(q)}$. The identifications
\begin{equation}
 (q,\psi)\sim\hat t_\ell(q,\psi)
 =\big(q+\ell,\,e^{2\pi i f_\ell(q)}\psi\big)
\end{equation}
therefore define a line bundle $L_{n,m}\to T^6$ with connection $d-2\pi iA$. Its first Chern class is represented by $dA=n\omega_1+m\omega_2$.

We now show that $L_{2,0}$ and $L_{0,4}$ descend to $X$. Under $r$ and $s$, the only nonzero shifts of $A$ are $s^*A-A=dy_1$ on $L_{2,0}$ and $r^*A-A=dy_3$ on $L_{0,4}$. We lift the actions of $r$ and $s$ on the base to the following maps $\hat r$ and $\hat s$ on the total space of each bundle:
\begin{equation}
\begin{array}{c|cc}
 & \hat r(q,\psi) & \hat s(q,\psi) \\ \hline
 L_{2,0} & (rq,\,i\psi) & (sq,\,e^{2\pi i y_1}\psi) \\
 L_{0,4} & (rq,\,e^{2\pi i y_3}\psi) & (sq,\,\psi)
\end{array}
\end{equation}
These maps preserve the connection and respect the lattice identifications.
The lifts satisfy the defining relations of $\Gamma$ modulo the
lattice identifications $\hat t_\ell$. 
For example, on $L_{2,0}$ one finds $\hat s^{\,2}=\hat t_{(1,0,0,0,0,0)}$ and $(\hat r\hat s)^4=\hat t_{(1,1,-1,-1,0,0)}$. 
Both bundles therefore descend to $X$, with first Chern classes represented by ${\cal D}_1$ and ${\cal D}_2$, respectively. Together with the periods on $C_1$ and $C_2$, this proves that ${\cal D}_1,{\cal D}_2$ form an integral basis of $H^2(X,\mathbb Z)$ modulo torsion.

\paragraph{Degree four.}  Up to permutations, the only nonzero triple intersection is $\kappa_{112}=\frac1{16}\int_{T^6}{\cal D}_1^2\wedge{\cal D}_2=2$. By Poincar\'e duality, the dual integral basis of $H^4(X,\mathbb Z)$ modulo torsion is $\widetilde{\cal D}^1=\frac12{\cal D}_1\wedge{\cal D}_2$ and $\widetilde{\cal D}^2=\frac12{\cal D}_1^2$, with $\int_X{\cal D}_a\wedge\widetilde{\cal D}^b=\delta_a^b$.

\subsection*{C. Primitive integral basis in degree three}

In the notation of the main text, the generators of $\Gamma$ act on the one-forms of $T^6$ as
\begin{equation}
\begin{aligned}
 r^*(e_1^\pm,e_2^\pm,e_3^\pm)&=(e_2^\pm,-e_1^\pm,e_3^\pm),\\
 s^*(e_1^\pm,e_2^\pm,e_3^\pm)&=(e_1^\pm,-e_2^\pm,-e_3^\pm).
\end{aligned}
\end{equation}
The six forms ${\cal A}_I,{\cal B}^I$ are therefore $\Gamma$-invariant.
Using $e_i^+\wedge e_i^-=2\,dx_i\wedge dy_i$ on each factor and
integrating over the covering torus gives
\begin{equation}
\begin{aligned}
 \int_X{\cal A}_I\wedge{\cal B}^J&=\delta_I^J,\\
 \int_X{\cal A}_I\wedge{\cal A}_J
 =\int_X{\cal B}^I\wedge{\cal B}^J&=0 .
\end{aligned}
\label{sm:symplectic}
\end{equation}
Since $b_3(X)=6$, these forms give a symplectic basis of $H^3(X,\mathbb R)$.

\paragraph{Integrality and primitivity.}
Consider the three-tori $S_i\subset T^6$ and the subgroups $K_i \subset \Gamma$ preserving them,
with orders in parentheses:
\begin{equation}
\begin{array}{c|c|c}
 i & S_i(t_1,t_2,t_3) & K_i \\ \hline
 1 & (t_1,t_2,t_2,t_1,t_3,0) & \langle r^2,rs\rangle\ (8)\\
 2 & (t_1,t_2,t_2,t_1,\tfrac18,t_3) & \langle rs\rangle\ (4)\\
 3 & (t_1,0,0,t_2,t_3,0) & \langle r^2,s\rangle\ (4)\\
 4 & (t_1,0,0,t_2,0,t_3) & \langle s\rangle\ (2)\\
 5 & (0,t_1,0,t_2,t_3,0) & \langle r\rangle\ (4)\\
 6 & (\tfrac14,t_1,\tfrac14,t_2,\tfrac14,t_3) & \langle r^2s\rangle\ (2)
\end{array}
\end{equation}
Each $K_i$ acts freely and preserves the orientation of $S_i$.
Thus $S_i/K_i$ defines an integral three-cycle $C_i$ in $X$.
For $(\eta_1,\ldots,\eta_6)=({\cal A}_0,{\cal A}_1,{\cal A}_2,
{\cal B}^0,{\cal B}^1,{\cal B}^2)$, the periods are
\begin{equation}
 P_{ai}=\int_{C_i}\eta_a=\frac{1}{|K_i|}\int_{S_i}\eta_a .
\end{equation}
Pulling back the forms to the tori in the table gives
\begin{equation}
 P=\begin{pmatrix}
 0&0&0&0&1&1\\
 0&0&-1&-1&0&0\\
 0&-1&0&0&0&1\\
 0&0&0&0&0&-1\\
 0&0&0&-1&0&0\\
 1&1&0&0&-1&-1
 \end{pmatrix}, 
\end{equation}
such that $\det P=1$. 
Since $P^{-1}$ has integer entries, the cycles $C'_b=\sum_i(P^{-1})_{ib}C_i$ are integral and satisfy $\int_{C'_b}\eta_a=(PP^{-1})_{ab}=\delta_{ab}$. By \eqref{sm:symplectic}, their Poincar\'e duals are ${\cal B}^0,{\cal B}^1,{\cal B}^2,-{\cal A}_0,-{\cal A}_1,-{\cal A}_2$, so all these six forms represent integral classes. For any integral class $[\alpha]=\sum_a u_a[\eta_a]$, we have $u_a=\int_{C'_a}\alpha\in\mathbb Z$. Since the period over $C'_a$ recovers the coefficient $u_a$, every integral class modulo torsion lies in the integer span of the $[\eta_a]$. Thus $({\cal A}_I,{\cal B}^I)$ is a primitive symplectic basis of $H^3(X,\mathbb Z)$ modulo torsion.

\subsection*{D. The O6 locus}

From the action of $\sigma$, one finds $\sigma^*{\cal D}_a=-{\cal D}_a$ and $\sigma^*e^\pm_i=\pm e^\pm_i$, so the ${\cal A}_I$ are even and the ${\cal B}^I$ odd. A point $[z]\in X$ is fixed by $\sigma$ if and only if $\sigma z=gz$ on $T^6$ for some $g\in\Gamma$, i.e.\ if $z$ is fixed by $g^{-1}\sigma$. We take $c\in\{0,1\}$ and write all coordinate equations modulo one. There are no solutions for the following eight elements:
\begin{itemize}
\item For $g=r^aw^c$ with $a$ odd, the third-torus equations require both $y_3=a/4$ and $2y_3=0$.
\item For $g=sw^c$ or $g=r^2sw^c$, the first-torus equations require $x_1+y_1$ or $x_1-y_1$, respectively, to take two values differing by $\tfrac12$.
\end{itemize}
For each remaining element, the solutions form a connected three-torus:
\begin{equation*}
\begin{array}{c|l}
 g & {\rm Fix}_{T^6}(g^{-1}\sigma) \\ \hline
 w^c & x_1+y_1=x_2+y_2=\frac c2, y_3=0\\
 r^2w^c & x_1-y_1=x_2-y_2=\frac c2, y_3=\frac12\\
 rsw^c & x_2=y_1+\frac c2, y_2=x_1+\frac{1+c}{2}, 2x_3+y_3=\frac14\\
 r^3sw^c & x_2=-y_1+\frac c2, y_2=-x_1+\frac{1+c}{2}, 2x_3+y_3=\frac34
\end{array}
\end{equation*}
These tori are disjoint, since a common point for $g\neq h$ would be fixed by $h^{-1}g\neq1$, whereas $\Gamma$ acts freely. They form three $\Gamma$-orbits of sizes $2,2,4$. As representatives we take the fixed loci of $\sigma$, $r^2\sigma$ and $(rs)^{-1}\sigma$, respectively,
\begin{equation}
\begin{aligned}
 S_A&=(t_1,-t_1,t_2,-t_2,t_3,0),\\
 S_B&=\big(t_1,t_1,t_2,t_2,t_3,\tfrac12\big),\\
 S_C&=\big(t_1,t_2,t_2,t_1+\tfrac12,\tfrac18+t_3,-2t_3\big).
\end{aligned}
\end{equation}
Their stabilizers are $\langle r,w\rangle$, $\langle r,w\rangle$ and $\langle r^2,w\rangle$, of orders $8,8,4$, respectively. Thus $\Sigma_{\rm O6}=\Sigma_A\sqcup\Sigma_B\sqcup\Sigma_C$, with $\Sigma_\alpha=S_\alpha/{\rm Stab}_\Gamma(S_\alpha)$, consists of three disjoint, embedded, flat three-manifolds. Each is special Lagrangian, since $J$ and ${\rm Im}\,\Omega$ are odd under $\sigma$ and hence restrict to zero on its fixed locus.

\paragraph{Orientations, periods and Poincar\'e dual.}
We orient each component by ${\rm Re}\,\Omega>0$. On $S_A$, $S_B$ and $S_C$, the pullback of ${\rm Re}\,\Omega$ is $dt_1\wedge dt_2\wedge dt_3$ times $2(1-\cos\theta)$, $-2(1+\cos\theta)$ and $4\nu\sin\theta$, respectively. The parameter orientation agrees with that defined by ${\rm Re}\,\Omega$ on $S_A$ and $S_C$. On $S_B$ the coefficient is negative, so we reverse the parameter orientation when computing its periods. Dividing by the stabilizer orders gives the periods
\begin{equation}
\begin{array}{c|c|rrr|rrr}
 & {\rm Re}\,\Omega & {\cal A}_0 & {\cal A}_1 & {\cal A}_2 & {\cal B}^0 & {\cal B}^1 & {\cal B}^2\\ \hline
 \Sigma_A & \frac{1-\cos\theta}{4} & 1 & 1 & 0 & 0 & 0 & 0\\
 \Sigma_B & \frac{1+\cos\theta}{4} & -1 & 1 & 0 & 0 & 0 & 0\\
 \Sigma_C & \nu\sin\theta & 0 & 0 & 2 & 0 & 0 & 0
\end{array}
\end{equation}
The matrix of ${\cal A}_I$ periods has determinant $4$, so the three components represent linearly independent classes in $H_3(X,\mathbb Q)$. With the convention $\int_\Sigma\zeta=\int_X\zeta\wedge{\rm PD}[\Sigma]$, the periods and \eqref{sm:symplectic} give, modulo torsion,
\begin{equation}
 {\rm PD}[\Sigma_{\rm O6}]=2\,({\cal B}^1+{\cal B}^2).
\end{equation}
As a check, $\int_X{\rm Re}\,\Omega\wedge{\rm PD}[\Sigma_{\rm O6}]=\frac12+\nu\sin\theta$ equals the sum of the three ${\rm Re}\,\Omega$ periods.


\begin{thebibliography}{99}


\bibitem{Coudarchet:2023mfs}
T.~Coudarchet,
``Hiding the extra dimensions: A review on scale separation in string theory,''
Phys. Rept. \textbf{1064} (2024), 1-28
[arXiv:2311.12105 [hep-th]].


\bibitem{Gautason:2015tig}
F.~F.~Gautason, M.~Schillo, T.~Van Riet and M.~Williams,
``Remarks on scale separation in flux vacua,''
JHEP \textbf{03} (2016), 061
[arXiv:1512.00457 [hep-th]].

\bibitem{Lust:2019zwm}
D.~L{\"u}st, E.~Palti and C.~Vafa,
``AdS and the Swampland,''
Phys. Lett. B \textbf{797} (2019), 134867
[arXiv:1906.05225 [hep-th]].

\bibitem{Cribiori:2022trc}
N.~Cribiori and G.~Dall'Agata,
``Weak gravity versus scale separation,''
JHEP \textbf{06} (2022), 006
[arXiv:2203.05559 [hep-th]].


\bibitem{DeWolfe:2005uu}
O.~DeWolfe, A.~Giryavets, S.~Kachru and W.~Taylor,
``Type IIA moduli stabilization,''
JHEP \textbf{07} (2005), 066
[arXiv:hep-th/0505160 [hep-th]].


\bibitem{Camara:2005dc}
P.~G.~Camara, A.~Font and L.~E.~Ibanez,
``Fluxes, moduli fixing and MSSM-like vacua in a simple IIA orientifold,''
JHEP \textbf{09} (2005), 013
[arXiv:hep-th/0506066 [hep-th]].


\bibitem{Farakos:2020phe}
F.~Farakos, G.~Tringas and T.~Van Riet,
``No-scale and scale-separated flux vacua from IIA on G2 orientifolds,''
Eur. Phys. J. C \textbf{80} (2020) no.7, 659
[arXiv:2005.05246 [hep-th]].


\bibitem{Cribiori:2021djm}
N.~Cribiori, D.~Junghans, V.~Van Hemelryck, T.~Van Riet and T.~Wrase,
``Scale-separated AdS4 vacua of IIA orientifolds and M-theory,''
Phys. Rev. D \textbf{104} (2021) no.12, 126014
[arXiv:2107.00019 [hep-th]].


\bibitem{Tringas:2025bwe}
G.~Tringas and T.~Wrase,
``Classical scale-separated AdS$_3$ vacua in heterotic string theory,''
[arXiv:2511.07781 [hep-th]].


\bibitem{Collins:2022nux}
T.~C.~Collins, D.~Jafferis, C.~Vafa, K.~Xu and S.~T.~Yau,
``On Upper Bounds in Dimension Gaps of CFT's,''
[arXiv:2201.03660 [hep-th]].


\bibitem{Bobev:2025yxp}
N.~Bobev, H.~Paul and F.~Revello,
``Holographic Constraint on Scale Separation,''
Phys. Rev. Lett. \textbf{136} (2026) no.18, 181601
[arXiv:2512.11031 [hep-th]].


\bibitem{Revello:2026yla}
F.~Revello and V.~Van Hemelryck,
``$\mathcal{N}=1$ spectra, cubic couplings and the rigid fate of DGKT,''
[arXiv:2607.00789 [hep-th]].


\bibitem{Catanese:2018}
F.~Catanese and A.~Demleitner,
\emph{The classification of Hyperelliptic threefolds},
arXiv:1812.09754 [math.AG] (2018).


\bibitem{Junghans:2020acz}
D.~Junghans,
``O-Plane Backreaction and Scale Separation in Type IIA Flux Vacua,''
Fortsch. Phys. \textbf{68} (2020) no.6, 2000040
[arXiv:2003.06274 [hep-th]].


\bibitem{Marchesano:2020qvg}
F.~Marchesano, E.~Palti, J.~Quirant and A.~Tomasiello,
``On supersymmetric AdS$_{4}$ orientifold vacua,''
JHEP \textbf{08} (2020), 087
[arXiv:2003.13578 [hep-th]].


\bibitem{Oguiso:1999xq}
K.~Oguiso and J.~Sakurai,
``Calabi-Yau threefolds of quotient type,''
Asian J. Math. \textbf{5} (2001) no.1, 43--77
[arXiv:math/9909175 [math.AG]].


\bibitem{Fischer:2013qza}
M.~Fischer, S.~Ramos-Sanchez and P.~K.~S.~Vaudrevange,
``Heterotic non-Abelian orbifolds,''
JHEP \textbf{07} (2013), 080
[arXiv:1304.7742 [hep-th]].




\bibitem{Fischer:2012qj}
M.~Fischer, M.~Ratz, J.~Torrado and P.~K.~S.~Vaudrevange,
``Classification of symmetric toroidal orbifolds,''
JHEP \textbf{01} (2013), 084
[arXiv:1209.3906 [hep-th]].


\bibitem{Braun:2011hd}
V.~Braun,
``The 24-Cell and Calabi-Yau Threefolds with Hodge Numbers (1,1),''
JHEP \textbf{05} (2012), 101
[arXiv:1102.4880 [hep-th]].


\bibitem{Derdzinski_2025}
A.~Derdzinski and P.~Piccione,
``Compact Flat Manifolds and Reducibility,''
in \emph{Advances in Submanifold Theory and Related Topics},
Springer Nature Switzerland (2025), pp.~181--208.


\bibitem{Doran:2014qna}
C.~Doran, S.~Mendez-Diez and J.~Rosenberg,
``String theory on elliptic curve orientifolds and KR-theory,''
Commun. Math. Phys. \textbf{335} (2015) no.2, 955-1001
[arXiv:1402.4885 [hep-th]].


\bibitem{Grimm:2004ua}
T.~W.~Grimm and J.~Louis,
``The Effective action of type IIA Calabi-Yau orientifolds,''
Nucl. Phys. B \textbf{718} (2005), 153-202
[arXiv:hep-th/0412277 [hep-th]].


\bibitem{Junghans:2023yue}
D.~Junghans,
``A note on O6 intersections in AdS flux vacua,''
JHEP \textbf{02} (2024), 126
[arXiv:2310.17695 [hep-th]].


\bibitem{Minasian:1997mm}
R.~Minasian and G.~W.~Moore,
``K theory and Ramond-Ramond charge,''
JHEP \textbf{11} (1997), 002
[arXiv:hep-th/9710230 [hep-th]].

\bibitem{Witten:1998cd}
E.~Witten,
``D-branes and K-theory,''
JHEP \textbf{12} (1998), 019
[arXiv:hep-th/9810188 [hep-th]].

\bibitem{Witten:1996md}
E.~Witten,
``On flux quantization in $M$-theory and the effective action,''
J. Geom. Phys. \textbf{22} (1997), 1-13
[arXiv:hep-th/9609122 [hep-th]].

\bibitem{Bergman:2001rp}
O.~Bergman, E.~G.~Gimon and S.~Sugimoto,
``Orientifolds, RR torsion, and K theory,''
JHEP \textbf{05} (2001), 047
[arXiv:hep-th/0103183 [hep-th]].

\bibitem{Marolf:2000cb}
D.~Marolf,
``Chern-Simons terms and the three notions of charge,''
[arXiv:hep-th/0006117 [hep-th]].




\bibitem{Lust:2004ig}
D.~Lust and D.~Tsimpis,
``Supersymmetric AdS(4) compactifications of IIA supergravity,''
JHEP \textbf{02} (2005), 027
[arXiv:hep-th/0412250 [hep-th]].


\bibitem{Plauschinn:2022ztd}
E.~Plauschinn,
``Mass spectrum of type IIB flux compactifications {\textemdash} comments on AdS vacua and conformal dimensions,''
JHEP \textbf{02} (2023), 257
[arXiv:2210.04528 [hep-th]].



\bibitem{Ooguri:2006in}
H.~Ooguri and C.~Vafa,
``On the Geometry of the String Landscape and the Swampland,''
Nucl. Phys. B \textbf{766} (2007), 21-33
[arXiv:hep-th/0605264 [hep-th]].



\bibitem{AAAZ}
A.~Aggarwal and A.~Zarafonitis, {\it Work in Progress}. 


\bibitem{Revello:2026eqp}
F.~Revello and V.~Van Hemelryck,
``Broken and restored: a holographic constraint for AdS vacua with orbifolds,''
[arXiv:2605.07464 [hep-th]].





\end{thebibliography}
\end{document}